\documentclass[prl,twocolumn,showpacs,showkeys,amssymb]{revtex4}
\usepackage{graphicx}

\begin{document}

\title{Dependence of ground state energy of classical $n$-vector spins on $n$} 

\author{Samarth Chandra}
\email{schandra@tifr.res.in}
\address{Department of Theoretical Physics, Tata Institute of Fundamental Research, Homi Bhabha Road, Colaba, Mumbai-400005, INDIA}

\date{\today}
\begin{abstract}

  We study the ground state energy $E_G(n)$ of $N$ classical $O(n)$ vector spins with the Hamiltonian $ \mathcal H = -\sum_{i>j} J_{ij} \vec S_i .\vec S_j  $ where the coupling constants \{$J_{ij}$\} are arbitrary. We prove that $E_G(n)$ is independent of $n$  for all $ n > n_{max}(N)  = \big \lfloor \frac{\sqrt{8N+1}-1}{2} \big \rfloor  $. We show that this bound is the best possible. We also derive an upper bound for $E_G(m)$ in terms of $E_G(n)$, for $m<n$. We obtain an upper bound on the frustration in the system, as measured by $F(n)  \equiv \frac{\sum_{i>j} |J_{ij}| + E_G(n)}{\sum_{i>j} |J_{ij}|} $. We describe a procedure for constructing a set of $J_{ij}$'s such that an arbitrary given state, \{$\vec{S}_i$\}, is the ground state.

\end{abstract}



\maketitle

\section{I. Introduction}

 In this paper, we study the ground states of $N$ unit classical $O(n)$ spins,  $\vec S_i$, having a hamiltonian of the form 
\begin{eqnarray}
 \mathcal H = -\sum_{i>j} J_{ij} \vec S_i .\vec S_j  
\end{eqnarray}
where $J_{ij}$'s are arbitrary real numbers--positive, negative or zero. Such hamiltonians with arbitrary bonds and couplings are of interest in the context of disordered systems, especially spin glasses [1]. One of the interesting questions is the behavior of the ground state energy as the spin space dimension, $n$, is increased. For example one can study the behavior of such models when $n$ is large. In this context, Hastings [2] proved that for $N$ spins beyond a spin space dimension of $ n_{max}(N)  = \big \lfloor \frac{\sqrt{8N+1}-1}{2} \big \rfloor $ the ground state energy does not decrease any further and also that this bound is saturated. Aspelmeier and Moore [3] have then used this bound in accelerating their numerical simulations of spin glasses. We provide an alternate proof for this bound. A similar analysis has been done earlier in the context of correlation matrices by Grone, et. al. [4] and for a relaxed version of maxcut problem of theoretical computer science [5].

 An interesting question is the behavior of the \emph{average} number of non-zero spin space components (average over disorder) in the ground state, as a function of the number of spins, $N$. For the infinite range model with gaussian distributed $J_{ij}$'s, this number increases as $N^{\mu}$ where $\mu = \frac{2}{5}$ [2,3]. Lee, Dhar and Young [6] have numerically determined $\mu$ for several different models. 

 We also derive both upper and lower bounds on the ground state energy of $O(m)$ spins in terms of the ground state energy when they are replaced by $O(n)$ spins ($m \neq n$) keeping the couplings, $J_{ij}$'s, the same. A stronger bound is also provided for Ising spins ($m=1$) when all couplings are antiferromagnetic and $E_G(n)$ is low.

 We also consider the problem of finding the ground state of such a hamiltonian [7]. We study the inverse problem---how to find a (non-trivial) hamiltonian of the form in equation (1) so that a given spin state \{$ \vec S_i$\} is the ground state. This question is trivial for Ising spins. One just assigns a non-negative $J_{ij}$ if the spins are parallel, and a non-positive $J_{ij}$ if they are anti-parallel. However with $O(n)$ spins $(n>1)$ the problem is non-trivial and in some cases there is no (non-zero) solution, for example, $N=3$, $n \geq 3$, with the three spins non-coplanar. In general, to find the desired set of couplings, $J_{ij}$'s, we can express the hamiltonian in terms of the angles of the spherical polar coordinates of the spins and set the derivatives with respect to the angles equal to zero at the angles corresponding to the desired ground state. This gives a set of linear relations between the couplings, $J_{ij}$'s. In addition, to ensure that this extremum is a minimum, and not a maximum or a saddle point, we have the additional constraint of the Hessian being positive semi-definite. Finding $J_{ij}$'s which simultaneously satisfy the linear relations as well as the positive semidefiniteness constraint on the Hessian is non-trivial. It is a semi-definite programming problem [8] for which fast algorithms and their software implementations are available.

 We provide a simple procedure for obtaining a large class of such hamiltonians. However, not all hamiltonians with \{$ \vec S_i$\} as the ground state are obtained by this procedure. We conjecture a characterisation of the hamiltonians obtained and give proof of a part of the conjecture.

 The plan of the paper is as follows. In section II we summarise some properties of the correlation matrices of classical spin states which are found useful in the later sections. In section III, we prove that for $N$ spins beyond a spin space dimension of $n_{max}(N)$ the ground state energy becomes independent of $n$. That this is the best bound is proved by providing a sequence of graphs and couplings, one for each $N$, such that $E_G(n_{max}-1) > E_G(n_{max})$. In section IV, we derive both upper and lower bounds on the ground state energy of $O(m)$ spins in terms of the ground state energy when they are replaced by $O(n)$ spins ($m \neq n$) keeping the couplings, $J_{ij}$'s, the same. For Ising spins with all couplings antiferromagnetic, in a special case, a stronger bound is derived. We obtain an upper bound on spin frustration, as measured by $F(n)  \equiv \frac{\sum_{i>j} |J_{ij}| + E_G(n)}{\sum_{i>j} |J_{ij}|} $ --- we show that $F(n) - F(\infty) \leq \beta_n$ where $\beta_n $ is a constant, independent of $J_{ij}$'s. In section V, we provide a procedure for constructing hamiltonians of the form in equation (1) with arbitrary given state \{$\vec {S}_i$\} as the ground state. Section VI summarises the results.

%%%%%%%%%%%%%%%%%%%%%%%%%%%%%%%%%%%%%%%%%%%%%%%%%%%%%%%%%%%%%%%%%%%%%%%%%%%%%%%%%%%%%%%%%%%%

\section{II. Some properties of correlation matrices of classical spin states}

 For an arbitrary state \{$\vec{S_i}$\} of $O(n)$ spins define the $(N \times N)$ correlation matrix, $C = [\vec S_i .\vec S_j]$.  Alternatively, 

\begin{equation}
C = S^T S
\end{equation}
where $S$ is the $(n \times N)$ matrix with vector of the $i^{th}$ spin as the  $i^{th}$ column. Clearly, $C$ is real, symmetric, has diagonal elements unity and can be written as $C = ODO^T$, where $O$ is an orthogonal matrix and $D$ diagonal.

 $C$ is positive semidefinite, i.e. all eigenvalues of $C$ are non-negative, since for every $x \in R^N, x^TCx = (Sx)^T(Sx) \geq 0 $.

 The number of non-zero (and hence positive) eigenvalues of $C$ is atmost $n$. This can be seen as follows: each row of $C$ is a linear combination of the $n$ rows of $S$ implying that the number of linearly independent rows of $C$ is atmost $n$. Diagonalising $C$, let $C = ODO^T$ where $O$ is an orthogonal matrix and $D$ diagonal with (let's say) first $k$ eigenvalues positive and rest zero. The rows (columns) of $O$ are mutually orthogonal and hence linearly independent. $(DO^T)$ now has $k$ linearly independent rows and thus $ODO^T$ also has $k$ linearly independent rows. The number of linearly independent rows of $C$ we have already argued to be atmost $n$. Hence the number of positive eigenvalues of $C$ is atmost $n$.

 Conversely, if $C$ is a real, symmetric matrix with diagonal elements unity and having $n$ or fewer positive eigenvalues and rest zero, then there exists a spin state of classical $O(n)$ unit spins for which it is the correlation matrix. To see this, $C$ being real and symmetric, can be diagonalised as $C = ODO^T = (O \sqrt{D})(O \sqrt{D})^T$ where $O$ is an orthogonal matrix and $D$ diagonal with first $k (\leq n)$ diagonal entries positive. The last $(N-n)$ rows of $(O \sqrt{D})^T$ are known to be zero and we drop them to define an $(n \times N)$ matrix $S$ such that $C=S^TS$. Since $c_{ii} = 1 \mbox{  } \forall \mbox{ } i$ each column of $S$ can be interpreted as a unit classical $O(n)$ spin, $C$ being their correlation matrix.

%%%%%%%%%%%%%%%%%%%%%%%%%%%%%%%%%%%%%%%%%%%%%%%%%%%%%%%%%%%%%%%%%%%%%%%%%%%%%%%%%%%%%%%%%%%%%%%%%

\section{III. The independence of the ground state energy from $n$ for $n \geq n_{max}(N)$}

 Consider the variation of the ground state energy $E_G(n)$ as a function of the spin space dimension $n$ of the $O(n)$ spins keeping the couplings, $J_{ij}$'s, the same. For $n' > n$  we have $E_G(n) \geq E_G(n')$ because for any state of $O(n)$ spins we can construct a corresponding state of $O(n')$ spins with the same value of energy by augmenting each vector with $(n' - n)$ zeroes. Also for any $n > N$ we have $E_G(n) = E_G(N)$ because $N$ spins span an atmost $N$ dimensional subspace of the $n$ dimensional spin space implying that by an appropriate choice of basis we can make all coordinates after the first $N$ coordinates zero and by dropping them we get an $O(N)$ spin state with the same value of energy.

\textbf{Theorem 1} For $N$ classical unit $O(n)$ vector spins with hamiltonian $\mathcal H = -\sum_{i>j} J_{ij} \vec S_i .\vec S_j$, where \{$J_{ij}$\} are any real numbers, the ground state energy $E_G(n) = E_G(n_{max})$ for all $n > n_{max}$ where 
\begin{equation}
 n_{max}(N) = \big \lfloor \frac{\sqrt{8N+1}-1}{2} \big \rfloor 
\end{equation}

 Here $\lfloor x \rfloor $ for $ x \in \mathbf R $ is the greatest integer not greater than $x$.

\textbf{Proof} Let us summarise the idea of the proof before getting into the details. Suppose we have a ground state which has more than $n_{max}(N)$ dimensions. Starting fron the correlation matrix of this state we construct another matrix which is the correlation matrix of a spin state which is embedded in one less spin space dimension but has the same energy. This construction always works whenever the spin state is embedded in more than $n_{max}(N)$ dimensions. Since, as shown above, $E_G(n) \geq E_G(n')$ for $n' > n$ this implies that for $n > n_{max}$, $E_G(n) = E_G(n_{max})$, as desired.

 We now discuss the proof. When the spins are $O(N)$ vectors let \{$\vec{S}_i$\} be a ground state. Consider the correlation matrix, $C$, with elements $ c_{ij} =  \vec S_i .\vec S_j $ for all $i$, $j$. Diagonalising $C$, we can write $C = O^T D O $ with $D$ a diagonal matrix and $O$ an orthogonal matrix. Let 

\begin{equation}
 D =  \left[ \begin{array}{cccccc}
d_1 & \ldots  & 0 & 0  & \ldots  & 0 \\
\ldots &     & \ldots  & 0 & \ldots & 0 \\
0 & \ldots & d_k & 0 & \ldots   & 0 \\
0 & \ldots & 0 &0 & & 0\\
 \ldots &  \ldots  & \ldots & &   &  \ldots  \\
0 & \ldots &  \ldots  & & & 0  \end{array} \right] 
\end{equation}

 where the first $k$ diagonal entries of $D$ are positive and rest zero.

 Consider $C' = O^T (D + rB)O$ where $B$ is symmetric with $B_{ij} = 0  \mbox{ if } i > k \mbox{ or } j > k$. This leaves $  \frac {1}{2} k (k+1) $ free parameters in $B$ and ensures that $C'$ is also symmetric and the zero eigenvalues of $C$ and the corresponding eigenvectors are not perturbed. Also, let $B$ satisfy

\begin{equation}
\label{Bdiag1}
  [O^T B O]_{tt} = 0
\end{equation}
for all $t=1,2, \ldots N$. This ensures that the diagonal elements of $C'$ remain unchanged.

 The $  \frac {1}{2} k (k+1) $ free parameters of $B$ must satisfy the $N$ linear homogenous equations (5). Hence whenever $ \frac {1}{2} k (k+1) > N $ such a non-zero $B$ will exist and we can increase $r$ till one of the first $k$ eigenvalues of $C$ becomes zero. Thus we obtain a matrix $C'$ which is the correlation matrix of a spin state embedded in $(k-1)$ dimensions.  As shown in the next paragraph, this spin state is a ground state. Thus applying this procedure repeatedly we obtain a ground state embedded in atmost $ \big \lfloor \frac{\sqrt{8N+1}-1}{2} \big \rfloor $ dimensions.

 The matrix $B$ chosen above is such that for $r$ small enough $C' =  C  \pm r O^T BO$ are \emph{both} correlation matrices of valid spin states with energy $ - \sum_{i>j} J_{ij} c_{ij} \pm r [\sum_{i>j} J_{ij}(O^TBO)_{ij}] $. Since we started from a $C$ which was a ground state this can happen only if $\sum_{i>j} J_{ij}(O^TBO)_{ij} = 0$ i.e. if $O^TBO$ was a neutral direction. Thus the correlation matrix $C'$ also corresponds to a ground state. 

 Hence we have provided a construction for continuously deforming a ground state and bringing it to lie in an atmost $n_{max}(N)$ dimensional subspace of the spin space without changing the energy, thus proving the desired result.

\textbf{Theorem 2} The bound in theorem 1 is the best possible, i.e. there exist values of \{$J_{ij}$\} such that $E_G(n_{max}) < E_G(n_{max}-1)$ where $ n_{max}(N) = \big \lfloor \frac{\sqrt{8N+1}-1}{2} \big \rfloor$.

\textbf{Proof} Consider three spins $\vec S_p$, $\vec S_q$ and $\vec S_{pq}$ with $J_{pq} = -J$ and $J_{p(pq)} = J_{q(pq)} = \sqrt{2} J $, see figure 1. It is easy to see that for this system of three spins in the ground state $\vec S_p$ is perpendicular to $\vec S_q$. If we integrate over $\vec S_{pq}$, we get an effective interaction between $\vec S_p$ and $\vec S_q$. Now construct a set of $k$ spins with this effective interaction between every pair of them (including the intermediate spins the total number of spins will be $N = \frac{1}{2} k (k+1)$). Their energy gets fully minimised only when these $k$ spins are perpendicular to each other which happens only in an atleast $n_{max} = k$ dimensional space thus completing the proof.

\begin{figure}
  \centering
  \includegraphics[width=0.9\columnwidth,angle=0]{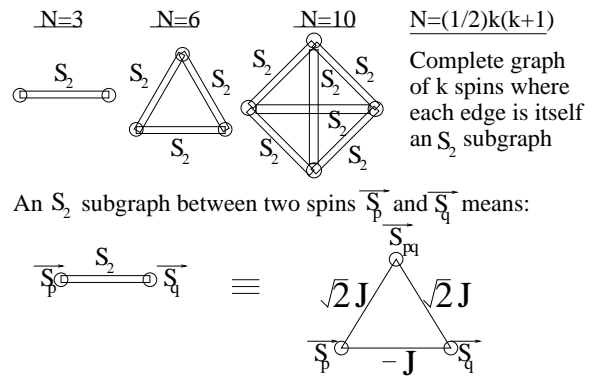}
  \caption{A sequence of examples for which $E_G(n_{max}(N)) < E_G(n_{max}(N)-1) $. Shown are three of those members of the sequence for which $ \frac{\sqrt{8N+1}-1}{2} $ is an integer (such $N$'s are $3,6,10,\ldots=\frac{1}{2}k(k+1)$). Rest of the members are obtained by adding appropriate number of free spins to the example of the last $N$ for which $\frac{\sqrt{8N+1}-1}{2}$ is an integer eg. $N=8$ example has two more free spins added to the $N=6$ example.}
  \label{advcorr}

\end{figure}

%%%%%%%%%%%%%%%%%%%%%%%%%%%%%%%%%%%%%%%%%%%%%%%%%%%%%%%%%%%%%

\section{IV. Bounds on the ground state energy}

 We have seen that $E_G(m) \geq E_G(n)$ for $m<n$. Now we will derive an \emph{upper} bound on $E_G(m)$ in terms of $E_G(n)$. This result (theorem 3) generalises a known result on the performance of Goemans-Williamson algorithm for maxcut problem of theoretical computer science [5]. The result by Goemans and Williamson, when translated into statistical physics language, would correspond to the special case of $m=1$. Theorem 4 is a translation of a known result on maxcut problem into statistical physics language [5]. The connection between the problem of finding the ground states of Ising spins and maxcut problem has been known before[9].

 It will be helpful to summarise the overall strategy before getting into the details. Suppose the various possible orientations of $O(m)$ spins occur according to an arbitrary given probability distribution. Then the energy is also a random variable and the expected value of the energy will always be greater than or equal to the ground state energy, i.e. $E_G(m) \leq E[\mathcal{H}_m]$, where $E[\mathcal{H}_m]$ denotes the expected value of the energy of $O(m)$ spins. If we choose the probability distribution in such a way that we are able to bound $E[\mathcal{H}_m]$ in terms of $E_G(n)$ from above we would have obtained the desired result.

 Now we give the derivation in detail. First we define a randomised procedure for obtaining an $O(m)$ state, say \{$\vec{S}_{i(m)}$\}, from the ground state \{$\vec{S}'_{i(n)}$\} of $O(n)$ spins. In the spin space of $O(n)$ spins randomly choose an $m$-dimensional subspace and project all the spins onto it. Normalise the $O(m)$ vectors thus obtained. Clearly different $O(m)$ states are obtained by this procedure depending on which $m$-dimensional subspace was chosen for projection. The expectation value of the $O(m)$ energy is $E[\mathcal{H}_m] = - \sum_{i>j} J_{ij} E[\vec{S}_{i(m)} . \vec{S}_{j(m)}]$. Now if $\mathcal{P}_{mn}$ denotes a projection operator from $n$ to $m$ dimensions

\begin{equation}
  E[\vec{S}_{i(m)} . \vec{S}_{j(m)}] =  \int \frac {\mathcal{P}_{mn} \vec{S}'_{i(n)} . \mathcal{P}_{mn} \vec{S}'_{j(n)}}{|\mathcal{P}_{mn} \vec{S}_{i(n)}| \mbox{ } |\mathcal{P}_{mn} \vec{S}_{j(n)}|  } d{\mathcal{P}_{mn}} \equiv f_{mn}(\theta_{ij}) 
\end{equation}
where $\theta_{ij}$ is the angle between $\vec{S}'_{i(n)}$ and $\vec{S}'_{j(n)}$ and the integral is over all projection operators $\mathcal{P}_{mn}$ with equal measure. As an example, in spherical polar coordinates,

\begin{widetext}
\begin{equation}
 f_{23}(\theta) = \int_{\phi_2 = 0}^{\pi} \int_{\phi_1 = 0}^{2 \pi} \frac{\sin\phi_2(\cos\theta - \cos\phi_1 \sin^2\phi_2 \cos(\phi_1 - \theta)  )}{4 \pi \sqrt{1-\cos^2\phi_1 \sin^2\phi_2} \sqrt{1- \sin^2\phi_2 \cos^2(\phi_1 - \theta)} }  \,d\phi_2 \,d\phi_1   
\end{equation}
\end{widetext}

By reversing the direction of $\vec{S}'_{i(n)}$ we observe that
\begin{equation}
 f_{mn}(\pi - \theta_{ij}) = - f_{mn}(\theta_{ij}) 
\end{equation}

Also $\frac {1-f_{mn}(\theta)}{1-\cos\theta} \geq 0 $ for all $\theta \in (0, \pi] $.
 Hence we can find a lower bound on $\frac {1-f_{mn}(\theta)}{1-\cos\theta}$, denoted by $\alpha_{mn}$, which gives
\begin{equation}
 f_{mn}(\theta) \leq (1 - \alpha_{mn}) + \alpha_{mn} \cos\theta 
\end{equation}

Also replacing $\theta$ by $(\pi - \theta)$ in this inequality we get
\begin{equation}
 - f_{mn}(\theta) \leq (1 - \alpha_{mn}) - \alpha_{mn} \cos\theta 
\end{equation}

For $J_{ij} < 0$, using (9), we get
\begin{equation}
 -J_{ij} f_{mn}(\theta_{ij}) \leq - (1- \alpha_{mn}) J_{ij} - \alpha_{mn} J_{ij} \cos\theta_{ij} 
\end{equation}

For $J_{ij} > 0$, using (10), we get
\begin{equation}
 -J_{ij} f_{mn}(\theta_{ij}) \leq (1- \alpha_{mn}) J_{ij} - \alpha_{mn} J_{ij} \cos\theta_{ij}
\end{equation}

Summing (11) over all those $ij$-pairs for which $J_{ij} < 0$ and (12) over all those $ij$-pairs for which $J_{ij} > 0$ and adding we get
\begin{equation}
 E[\mathcal{H}_m] = - \sum_{i>j} J_{ij} f_{mn}(\theta_{ij}) \leq (1-\alpha_{mn}) \sum_{i>j} |J_{ij}| + \alpha_{mn} E_G(n) 
\end{equation}

Now the minimum value of a random variable is always less than or equal to its expectaion value.

Therefore, we have

\textbf{Theorem 3} For $(m < n)$
\begin{equation}
 E_G(n) \leq E_G(m) \leq (1- \alpha_{mn}) \sum_{i>j} |J_{ij}| + \alpha_{mn} E_G(n) 
\end{equation}
where $\alpha_{mn}$ is the minimum value of $\frac{1-f_{mn}(\theta)}{1-\cos\theta} $ over the interval $\theta \in (0, \pi] $ and $f_{mn}(\theta)$ has been defined above. 

Or rearranging the inequality, $\left(\frac{1}{\alpha_{mn}} \right) E_G(m) - \left( \frac{1-\alpha_{mn}}{\alpha_{mn}} \right) \sum_{i>j} |J_{ij}| \leq E_G(n) \leq E_G(m)$

 As an example, for $m=1$ and $n$ arbitrary, $f_{1n}(\theta_{ij}) = 1 - 2 \frac{\theta_{ij}}{\pi} $ and $\alpha_{1n} \approx 0.87856 $ [5]. We have determined $f_{23}$ and $f_{34}$ numerically by representing them as integrals in spherical polar coordinates, see equation (7) for instance. The graphs of $f_{34}(\theta)$ and $q_{34}(\theta) = \frac{1-f_{mn}(\theta)}{1-\cos\theta}$ are shown in figure 2. We find that $\alpha_{23} \approx 0.96$ and $\alpha_{34} \approx 0.98$.

\begin{figure}
  \centering
  \includegraphics[width=0.7\columnwidth,angle=270]{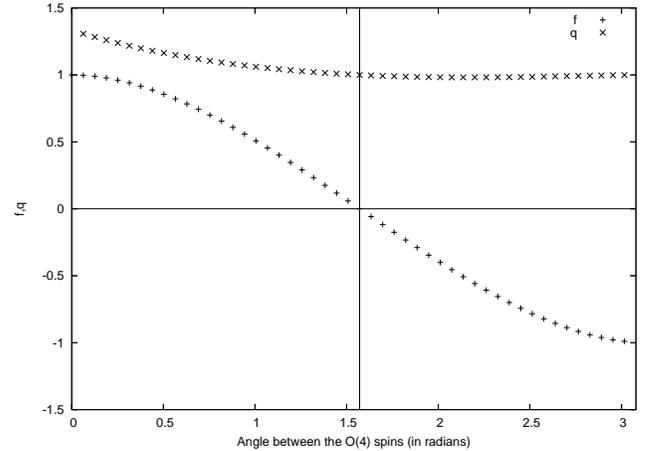}
  \caption{The functions $f_{34}(\theta)$ and $q_{34}(\theta) = \frac{1-f_{34}(\theta)}{1-\cos\theta}$  }
  \label{advcorr}

\end{figure}

 As a specific instance, for the triangular lattice anti-ferromagnet $E_G(1) = - \frac {1}{3} J$ implying that $ -(0.52)J \leq E_G(2) \leq -\frac{1}{3} J $. It is known that $E_G(2) =  -(0.5)J $ which compares very well with the \emph{non-trivial part} of the inequality.

 Now, we prove a stronger bound for a special case of Ising antiferromagnets:

\textbf{Theorem 4} For the special case of $m=1$, let all couplings, {$J_{ij}$}, be antiferromagnetic. Then for the case $(E_G(n) \leq \delta \sum_{i>j} |J_{ij}|)$ ($\delta \approx -0.69 $) we have the stronger bound

\begin{eqnarray}
& & E_G(n)  \\ & \leq & E_G(1) \\ & \leq & (-\sum_{i>j} |J_{ij}|) \frac{2}{\pi} \arccos \left( \frac{E_G(n)}{\sum_{i>j}|J_{ij}|} \right) + \sum_{i>j} |J_{ij}| 
\end{eqnarray}

\textbf{Proof} Again using the randomised procedure in the derivation of theorem 3, from the ground state \{$\vec{S}'_{i(n)}$\} of $O(n)$ model, various Ising states are obtained with different probabilities such that
$ E[\mathcal H_1] = - \sum_{i>j} J_{ij} + \sum_{i>j} J_{ij} \frac{2}{\pi} \arccos x_{ij} $
where $x_{ij} = \vec {S}'_{i(n)} . \vec {S}'_{j(n)} $ (using $f_{1n}(\theta_{ij}) = 1 - 2 \frac{\theta_{ij}}{\pi} $).

 Consider the function $\arccos x$. Draw the \emph{oblique} tangent from (1,0) to the curve, intersecting the curve tangentially at $(\delta, \arccos \delta )$. Consider the function $h(x)$ which is the same as $\arccos x$ for $x< \delta$ and the same as the tangent for $x \in [\delta, 1]$.
 
 Clearly, $ \arccos x_{ij} \geq  h(x_{ij}) $ and since all $J_{ij} \leq 0$, 

\begin{eqnarray}
& & E[\mathcal{H}_1]  \\  & \leq & \frac{2}{\pi} \sum_{i>j} J_{ij} h(x_{ij}) -\sum_{i>j} J_{ij} \\
& \leq & -\frac{2}{\pi} (\sum_{i>j} |J_{ij}|) h \left( \sum_{i>j} \frac{|J_{ij}|}{\sum_{p>q} |J_{pq}|} x_{ij} \right) - \sum_{i>j} J_{ij} 
\end{eqnarray}
 where the last inequality uses the convexity of $h(x)$.

 Since the minimum value of a random variable is less than or equal to its expectation value, we have $E_G(n) \leq E_G(1)  \leq  E[\mathcal H_1] $. Also, for $ x < \delta$ we have $h(x) = \arccos x$ and the desired inequality is proved.

 In the presence of antiferromagnetic $J_{ij}$'s, there may not exist any spin configuration that minimises the energy of each individual bond to $-|J_{ij}|$. One of the possible measures of the frustration of spins is 

\begin{equation}
 F(n) \equiv \frac{\sum_{i>j} |J_{ij}| + E_G(n)}{\sum_{i>j} |J_{ij}|} 
\end{equation}

 We can consider spin frustration as arising in two steps: first we choose the $J_{ij}$'s but do not put any restriction on the dimensionality of the spin space---it is allowed to be as large as desired for the minimisation of energy. The frustration of this system will be $F(\infty)$ which will be the same as $F(N)$ because $E_G(n) = E_G(N)$ for $n>N$. To obtain the actual $O(n)$ system we \emph{now} restrict the number of dimensions in the spin space to $n$, thus increasing the spin frustration from $F(\infty)$ to $F(n)$.

\textbf{Theorem 5} If $(m < n)$

\begin{equation}
 \frac{E_G(m) - E_G(n)}{\sum_{i>j}|J_{ij}|} \leq 2 (1-\alpha_{mn}) 
\end{equation}
As a particular case, 
\begin{equation}
 F(n)-F(\infty) \leq  2 (1-\alpha_{nN}) 
\end{equation}
where $\alpha_{mn}$ are the same as in theorem 3.

\textbf{Proof} In theorem 3, subtract $E_G(n)$ throughout, divide by $\sum_{i>j} |J_{ij}|$ and observe that $\frac{E_G(n)}{\sum_{i>j} |J_{ij}|} \geq -1$ thus completing the proof.

 In particular, for all $n$, $F(n) - F(\infty) \leq F(1) - F(\infty) \leq  2(1-\alpha_{1N}) \approx 0.24288 $.

\section{V. Procedure for constructing a model with an arbitrary given ground state}

 For $N$ classical spins of $O(n)$ type let \{$\vec{S}'_i$\} be a given state. We want to construct a hamiltonian with only two-spin Heisenberg type interactions which has \{$\vec{S}'_i$\} as the ground state. The following procedure constructs a hamiltonian of the form in eq. (1) (upto a constant) which has \{$\vec{S}'_i$\} as the ground state.

1. For the given ground state \{$\vec{S}'_i$\} construct the correlation matrix, $C'$, such that $c'_{ij} = \vec{S}'_i . \vec{S}'_j \mbox{  } \forall i,j = 1, 2, \ldots N$.

2. Let $C' = O' D' O'^T$ where $O'$ is an orthogonal matrix and $D'$ diagonal with, let's say, the first $k$ diagonal entries non-zero and rest entire matrix zero.

3. Construct an $(N \times N)$ auxilliary matrix $G$ as follows: 

\begin{equation}
  G = \left[ \begin{array}{cc}
  G_1 & G_2 \\
  G_3 & G_4 \end{array} \right] 
\end{equation}

where $G_1$ is a $(k \times k)$ matrix, etc. Moreover, choose $G_1 = 0$, $G_2 = 0$, $G_3 = 0$ and $G_4$ to be any $(N-k) \times (N-k)$ real, symmetric matrix with all eigen values \emph{non-positive}.

4. Define $J = [J_{ij}] = O' G O'^T $.

\textbf{Theorem 6}: For the hamiltonian 
$ \mathcal H = -\sum_{i,j = 1}^N J_{ij} \vec S_i .\vec S_j $
thus constructed, the spin state \{$\vec{S}'_i$\} is the ground state.

\textbf{Proof}: For any spin state \{$\vec{S}_i$\}, construct the correlation matrix $C= [c_{ij}] = [\vec{S}_i . \vec{S}_j]$ and diagonalise it, 
\begin{equation}
\label{Cdiag}
C = ODO^T
\end{equation}
where $O$ is an orthogonal matrix and $D$ is diagonal with all eigenvalues non-negative. 

 Also for the matrix $J$ defined above let
\begin{equation}
\label{Jdiag}
(-J)^T = \hat{O} \hat{D} \hat{O}^T
\end{equation}
 where $\hat{O}$ is an orthogonal matrix and $\hat{D}$ is diagonal. Since $J$ is negative semidefinite, $(-J)$ is positive semidefinite, thus the entries of $\hat{D}$ are non-negative.
 
 Now $\mathcal{H} = Tr((-J)^TC)$. Using equations (25) and (26) and repeatedly using $Tr(AB) = Tr(BA)$ we get $\mathcal{H} = Tr[W^T W] \geq 0$ where $W = \sqrt{\hat{D}} \hat{O}^T O \sqrt{D}$. Therefore, for any state \{$\vec{S}_i$\},
\begin{equation}
\label{Hpos}
\mathcal{H} \geq 0
\end{equation}

 For \{$\vec{S}'_i$\}, by construction, $\mathcal{H} = Tr((-J)^T C') = 0 $ implying that \{$\vec{S}'_i$\} is a ground state of $\mathcal{H}$.

%%%%%%%%%%%%%%%%%%%%%%%%%%%%%%%%%%%%%%%%%%%

 Although a large number of hamiltonians with arbitrary given state \{$\vec{S}'_i$\}  as the ground state can be obtained by this procedure, not all the hamiltonians with this property are obtained. For instance, it can be easily checked that for three Ising spins, one up and the other two down, happens to be a ground state when all three couplings are antiferromagnetic with equal strength, but this set of couplings can not be obtained by the above procedure for any allowed choice of the matrix $G_4$. Thus we would like to characterise which hamiltonians can be obtained by this procedure for a given ground state and which hamiltonians can not be obtained.

 We expect that a hamiltonian with \{$\vec{S}'_i$\} as the ground state is obtained by this procedure if and only if upon replacing the given spins by spins with any higher spin space dimension, keeping $J_{ij}$'s the same, the ground state energy remains the same. The \emph{if} part is our conjecture while the \emph{only if} part is proved as follows: for any $\tilde{n} > n$, by augmenting each vector of \{$\vec{S}'_i$\} by $(\tilde{n} - n)$ zeroes, we can obtain a state with the value of the hamiltonian $\mathcal H_{\tilde{n}} = 0$. Since the hamiltonian is expressible as the trace of the product of two symmetric positive semidefinite matrices its value cannot be negative as in equation (27), implying that the $\tilde{n}$ dimensional state thus obtained is the ground state of $O(\tilde{n})$ spins. Therefore, for $\tilde{n}>n$ we have $E_G(\tilde{n}) = E_G(n)$ thus completing the proof. This proof is consistent with the case of three Ising spins with antiferromagnetic couplings discussed above because if we replace three Ising spins by XY-spins the ground state energy decreases from $- \frac{1}{3} J $ per bond to $- \frac{1}{2} J $ per bond.

%%%%%%%%%%%%%%%%%%%%%%%%%%%%%%%%%%%%%%%%%%%%%%%%%

\section{VI. Summary}

 We showed that as we increase the spin space dimension, $n$, the ground state energy, $E_G(n)$, becomes independent of $n$ beyond a spin space dimension of $n_{max}(N)$, and this bound is the best possible. For $m<n$ we derived an upper bound for $E_G(m)$ in terms of $E_G(n)$, the lower bound was trivial. A stronger version for a special case of $m=1$ was also proved. Similar bounds on $ E_G(m) - E_G(n)$ and a measure of spin frustration, $F(n)$, were derived. A procedure was given for constructing a hamiltonian with an arbitrary given spin state, \{$\vec{S}'_i$\}, as the ground state.

 I want to thank Prof. Deepak Dhar for his guidance and encouragement throughtout this work. I thank Prof. Daya Gaur for introducing me to the area of approximation algorithms for NP-complete problems and maxcut problem. I thank Abhishek Dhar, Sriram Shastry, Alan Bray, Stephan Boyd and Katya Schienberg for discussions, Kanval Rekhi Foundation for partial financial support and CSIR for Shyama Prasad Mukherjee fellowship.

\end{document}